\documentclass[10pts,showpacs,preprint,aps]{revtex4}
\usepackage{graphicx}
\usepackage{dcolumn}
\usepackage{bm}
\usepackage{amssymb}
\usepackage{amsmath}
\begin{document}
\setcounter{page}{1}
\vskip 2cm
\title 
{About the propagation of the Gravitational Waves in an asymptotically de-Sitter space: Comparing two points of view}
\author
{Ivan Arraut}
\affiliation{  
Department of Physics, Osaka University, Toyonaka, Osaka 560-0043, Japan}

\begin{abstract}
We analyze the propagation of gravitational waves (GWs) in an asymptotically de-Sitter space by expanding the perturbation around Minkowski and introducing the effects of the Cosmological Constant ($\Lambda$), first as an additional source (de-Donder gauge) and after as a gauge effect ($\Lambda$-gauge). In both cases the inclusion of the Cosmological Constant $\Lambda$ impedes the detection of a gravitational wave at a distance larger than $L_{crit}=\left(6\sqrt{2}\pi f \hat{h}/\sqrt{5}\right)r_\Lambda^2$, where $r_\Lambda=\frac{1}{\sqrt{\Lambda}}$ and f and $\hat{h}$ are the frequency and strain of the wave respectively. We demonstrate that $L_{crit}$ is just a confirmation of the Cosmic No hair Conjecture (CNC) already explained in the literature.
\end{abstract}
\preprint{OU-HET 760-2012}
\maketitle 

\section{Introduction}

If we want to analyze the propagation of a gravitational wave (GW) in de Sitter space, we have two possibilities. The first one is to perform the appropriate expansion around the background \cite{7, 8K, S1, S2, S3}. And the second one is to perform the expansion around the flat Minkowskian background, but keeping in mind that the Cosmological Constant $\Lambda$ should be introduced as an additional source \cite{2} or as a gauge effect \cite{6}.\\
Many studies of GWs around the full de-Sitter background have already been performed. In \cite{8K} for example, the dynamical evolution for axisymmetric and non rotating GWs in asymptotically de-Sitter spacetime was analyzed.
In \cite{S1} the study of the geodesic deviations showed that $\Lambda$ enters as an isotropic motion inside the geodesic deviation equations. In \cite{S2, S3}, the simplicity of the Conformal transformations is used in order to analyze the behaviour of waves (electromagnetic or gravitational) around the de-Sitter background. In \cite{S4}, the cosmic no-hair conjecture (CNC) under the presence of GWs was analyzed by using analytical methods. In \cite{S5} the global structure of Robinson-Trautman radiative spacetimes with a positive $\Lambda$ was analyzed confirming the CNC under the presence of GWs.    
\\ 
If we analyze the GWs by expanding around Minkowski, then in the de-Donder gauge $\Lambda$ is an additional source and the full solutions, namely, plane wave plus $\Lambda$ contribution must satisfy the condition $\partial_\mu h^\mu_{\;\;\;\nu}=\frac{1}{2}\partial_\nu h$.\\
In the $\Lambda$ gauge condition, although $\Lambda$ is not anymore an explicit source \cite{6}, it appears in the final solution ($h_{\mu \nu}$) in order to satisfy the gauge condition $\partial_\mu h^\mu_{\;\;\;\nu}-\frac{1}{2}\partial_\nu h=\Lambda x_\nu$ (working with the appropriate signature as is given in \cite{2}).\\
In \cite{2} the de-Donder gauge was used and it was demonstrated that there exists a critical distance $(L_{crit})$ after which the GW cannot propagate anymore. It was however mentioned that such a distance is of astrophysical order of magnitude. In this paper we show that that is not the case if we take into account the standard background condition $r_\Lambda\approx \frac{\lambda}{h}$ already suggested by Misner, Thorne and Wheeler \cite{7}.\\ 
If we work in a different gauge ($\Lambda$ gauge), the physics is unchanged. Then we expect not only to find the appropriate behaviour of a spin-2 field under rotations of the system, but we also expect to find the same $L_{crit}$. This is the main motivation for this paper.
$L_{crit}$ is then a gauge independent quantity. It means that $L_{crit}$ is not only a coordinate effect, but a physically relevant quantity. $L_{crit}$ is simply related to the standard background condition $r_{\Lambda}\approx\frac{\lambda}{h}$ \cite{7}, where h is the strain of the wave and $\lambda$ is the wavelength for a monochromatic source. There is no loss of generality if we only analyze monochromatic sources since real waves are simply superposition of monochromatic plane ones.\\
Although $L_{crit}\approx r_\Lambda$ is just the standard background (global) condition \cite{7}, the present local analysis demonstrates that the background condition is just a consequence of a power decrease of the GW as it propagates inside the background \cite{2}. This is consistent with the analysis performed in \cite{8K, S4, S5} where the CNC was proved. 
The outline of the paper is as follows. In section \ref{eq:s1} we briefly review the results obtained in \cite{2} where we use the de-Donder gauge condition in order to obtain the first order solutions of the field equations and we obtain the Poynting vector for a monochromatic wave inside a de-Sitter background and the critical distance $L_{crit}$. In section \ref{eq:s3} we analyze the equations of motion in the $\Lambda$ gauge and the local coordinate transformations. In section \ref{eq:s5}, we demonstrate that we still have spin-2 behavior in the $\Lambda$ gauge with 2 physically relevant components. This is consistent with the results obtained in \cite{S1}. In section \ref{eq:s6} once again we calculate the power of a gravitational wave and the critical distance but this time in the $\Lambda$ gauge, showing then that the physics involved is gauge independent. In section \ref{eq:Stochastic}, we compare the results with those corresponding to the Stochastic gravitational waves background but focusing in the de-Sitter inflationary phase scenario.          
  
\section{Linearized Einstein's equations with $\Lambda$ and the Energy Momentum tensor. Lorenz gauge}   \label{eq:s1}

If we use the weak field approximation, the metric can be written as $g_{\mu \nu}=\eta_{\mu \nu}+h_{\mu \nu}$ \cite{1}, where $\eta_{\mu \nu}$ is the Minkowski metric. The field equations in the weak field approximation and with a positive $\Lambda$, after the imposition of the de-Donder gauge $\partial^\mu h_{\mu \nu}=\frac{1}{2}\partial_\nu h$ become \cite{2,6}:

\begin{equation}   \label{eq:5}
\square h_{\mu \nu}=-16\pi GS_{\mu \nu}-2\Lambda \eta_{\mu \nu}
\end{equation}

Where we ignore terms of order $\Lambda h$. The full solution for $h_{\mu \nu}$ can be expressed as:
  
\begin{eqnarray}   \label{eq:7}
h_{0 0}=e_{0 0}(\vec{r},\omega)e^{ikx}+c.c-\Lambda t^2\;\;\;\;\;h_{0 i}=e_{0 i}(\vec{r},\omega)e^{ikx}+c.c+\frac{2}{3}\Lambda tx_i\\\nonumber
h_{i j}=e_{i j}(\vec{r},\omega)e^{ikx}+c.c+\Lambda t^2\delta_{i j}+\frac{1}{3}\Lambda\epsilon_{i j}
\end{eqnarray} 
  
where $\epsilon_{i j}=x_i x_j$ for $i\neq j$ and 0 otherwise. The full solution of course respects the de-Donder condition $\partial^\mu h_{\mu \nu}=\frac{1}{2}\partial_\nu h$.
From the full solution, it can be observed that the particular one corresponds to the de-Sitter Background. This is the solution obtained for the graviton in the massless limit ($m\to0$) as can be shown from the results obtained in \cite{3}.
Additionally, in agreement with \cite{2}, the energy-momentum (tensor) carried by the gravitational waves with a positive $\Lambda$, is given by:  

\begin{equation}   \label{eq:9}
\hat{t}_{\mu \nu}=t_{\mu \nu}-\frac{1}{8\pi G}\Lambda h_{\mu \nu}
\end{equation}

If we expand up to second order and take into account that the first order Ricci scalar is given by $R^{(1)}=-\eta_{\mu \nu}\Lambda$, then the Poynting vector corresponding to the background solution becomes \cite{2}:

\begin{equation}   \label{eq:13}
\hat{t}_{0 i}=\frac{1}{8\pi G}\left(\frac{10}{9}\Lambda^2t x_i\right)
\end{equation}

If we now assume a wave moving along the z-direction, then the relevant quantity for us is (remember that after averaging the contributions from $h\Lambda\to0$ in agreement with \cite{2}):

\begin{equation}   \label{eq:14}
<t^{03}>=<t^{03}>_{wave}+<t^{03}>_\Lambda
\end{equation}   
   
Note, the subscript "wave" refers to the standard contribution without $\Lambda$. The critical distance \cite{2}, is obtained as $<t^{03}>=0$. After computation, it is given by:

\begin{equation}   \label{eq:15}
L_{crit}=\frac{6\sqrt{2}\pi f\hat{h}}{\sqrt{5}}r^2_\Lambda
\end{equation}

which depends on the frequency and the amplitude of the wave. $L_{crit}$ could in principle be of any order of magnitude. But it is just the background scale $r_\Lambda$ as can be seen after taking into account the condition $r_\Lambda\approx \frac{\lambda}{h}$ \cite{7}. The wave cannot propagate to a distance larger than $L_{crit}$, it means that the inhomogeneities (in this case GW) eventually disappear and the space goes asymptotically to the de-Sitter one. This is in agreement with deeper studies performed in \cite{8K, S4, S5} where the CNC in the presence of GWs was confirmed.  

\section{The $\Lambda$ gauge}   \label{eq:s3}

The $\Lambda$ gauge was already explained in \cite{6}. In such a gauge, $\Lambda$ is not a source anymore, but its effects can be seen as a gauge ones. In this gauge, we expect to have 2 physical polarization tensors and we expect to obtain the same $L_{crit}$ already found in eq. \ref{eq:15}. By proving this, we show that the background effects are gauge-independent as it must be.

\subsection{The equations of motion in the $\Lambda$ gauge}   \label{eq:s4}

The field equations in the $\Lambda$ gauge can be obtained if we introduce the gauge condition $\partial_\mu h^\mu_{\;\;\;\nu}-\frac{1}{2}\partial_\nu h=\Lambda x_\nu$ inside the full weak field version of the Einstein's equations given in \cite{2}. In this gauge, the equations simplify to:

\begin{equation}   \label{eq:16}
\square h_{\mu \nu}=0
\end{equation}

Where the solutions for this equation must be obtained after taking into account the gauge condition and were already found in \cite{6} and repeated here for clarity:

\begin{equation}   \label{eq:20}
h_{\mu \nu}=e_{\mu \nu}e^{ikx}+c.c+\frac{\Lambda}{18}\left(4x_\mu x_\nu-\eta_{\mu \nu}x^2\right)
\end{equation}   

The first part of this solution is the plane wave contribution. It is already known that under a general coordinate transformation, $h_{\mu \nu}$ transforms as shown in \cite{1}. The field equations (eqn. \ref{eq:16}) in momentum space show the fact the the graviton propagates at the light velocity $k^\mu k_\mu$. The gauge condition, on the other hand, can be written partially in momentum space and is given by:

\begin{equation}   \label{eq:19}
k_\mu e^\mu_{\;\;\nu}=\frac{1}{2}k_\nu e+\Lambda x_\nu
\end{equation}

If $h_{\mu \nu}$ does not satisfy the $\Lambda$ gauge condition, then it is always possible to find some $h^{'}_{\mu \nu}$ that does by performing the appropriate coordinate transformations with \cite{1}:

\begin{equation}   \label{eq:22}
\square\epsilon_\nu\equiv \partial_\mu h^\mu_{\;\;\;\nu}-\frac{1}{2}\partial_\nu h
\end{equation}   
 
if the $\Lambda$ gauge condition is satisfied, then the previous equation can only be reduced to:

\begin{equation}   \label{eq:23}
\square\epsilon_\nu=\Lambda x_\nu
\end{equation}  
 
the homogeneous solutions for these equations are the standard ones already given in \cite{1}:

\begin{equation}  \label{eq:24}
\epsilon^\mu(x)=i\epsilon^\mu e^{ikx}-i\epsilon^{\mu *}e^{-ikx}
\end{equation}

the particular solutions are:

$$
\begin{matrix}   \label{eq:251}
\epsilon^0_\Lambda=at^3+br^2t\\\nonumber
\epsilon^i_\Lambda=ct^2x^i+d(x^i)^3+ex^i((x^{j})^2+(x^{k})^2)
\end{matrix}
$$

the constants must satisfy the conditions $-a+b=\frac{\Lambda}{6}$ and $-c+6d+4e=\Lambda$ in agreement with equation (\ref{eq:23}). We will write these particular solutions as $\epsilon^\mu_\Lambda(x)$. Then, the infinitesimal parameters for the coordinate transformations can be written as:

\begin{equation}  \label{eq:25}
\epsilon^\mu(x)=i\epsilon^\mu e^{ikx}-i\epsilon^{\mu *}e^{-ikx}+\epsilon^\mu_\Lambda(x)
\end{equation}  

\section{Polarizations in the $\Lambda$ gauge}   \label{eq:s5}

If we assume that the wave is propagating along the z direction, namely, $k^1=k^2=0$ and $k^3=k^0=k>0$ \cite{1}; then the relations among the polarizations components can be obtained from eq. \ref{eq:19} as:

\begin{equation}   \label{eq:26}
e_{01}=-e_{3 1}+\left(\frac{\Lambda}{\omega}\right)x\;\;\;\; e_{02}=-e_{32}+\left(\frac{\Lambda}{\omega}\right)y\;\;\;\; e_{03}=-\frac{1}{2}(e_{33}+e_{00})+\left(\frac{\Lambda}{\omega}\right)z
\end{equation}

the previous equations correspond to the $\Lambda$ gauge condition written in the form \ref{eq:19} for $\nu=1,2,3$ respectively. For the case $\nu=0$, we get:

\begin{equation}   \label{eq:261}
e_{03}=-\frac{1}{2}(e_{33}+e_{00})-\left(\frac{\Lambda}{\omega}\right)t
\end{equation} 

if we sum this result with the last one obtained in \ref{eq:26}, we get:

\begin{equation}   \label{eq:262}
e_{03}=-\frac{1}{2}(e_{33}+e_{00})-\left(\frac{\Lambda}{\omega}\right)(z-t)
\end{equation} 

for a wave traveling along the z-direction, the relevant coordinates are $z$ and $t$. As the graviton must travel along a light cone, then the assumption $z=t$ is valid and the previous equations just become to be the same as the standard ones obtained in \cite{1}:

\begin{equation}   \label{eq:263}
e_{01}=-e_{3 1}\;\;\;\; e_{02}=-e_{32}\;\;\;\; e_{03}=-\frac{1}{2}(e_{33}+e_{00})
\end{equation}

if additionally, we analyze the transformations for the polarization tensors \cite{1}, then the independent components transform as:

\begin{equation}   \label{eq:27}
e^{'}_{11}=e_{11}-2\left(ct^2+3dx^2+e(y^2+z^2)\right)
\end{equation}

\begin{equation}   \label{eq:28}
e^{'}_{12}=e_{12}-4exy
\end{equation}

\begin{equation}   \label{eq:288}
e^{'}_{13}=e_{13}+k\epsilon_1-4exz
\end{equation}

\begin{equation}   \label{eq:29}
e^{'}_{23}=e_{23}+k\epsilon_2-4eyz
\end{equation}

\begin{equation}   \label{eq:30}
e^{'}_{33}=e_{33}+2k\epsilon_3-2\left(ct^2+3dz^2+e(y^2+x^2)\right)
\end{equation}

\begin{equation}   \label{eq:31}
e^{'}_{00}=e_{00}-2k\epsilon_0-2\left(3at^2+br^2\right)
\end{equation}

observing these expressions for the 6 independent polarization tensors, we might believe that there is a problem here because we are expecting only 2 of them to have physical relevance. However, if $z=t$ and if additionally we ignore the coordinates $x$ and $y$ since the wave is propagating along z, the previous transformations just become:

\begin{equation}   \label{eq:271}
e^{'}_{11}=e_{11}-2z^2\left(c+e\right)
\end{equation}

\begin{equation}   \label{eq:281}
e^{'}_{12}=e_{12}
\end{equation}

\begin{equation}   \label{eq:2881}
e^{'}_{13}=e_{13}+k\epsilon_1
\end{equation}

\begin{equation}   \label{eq:291}
e^{'}_{23}=e_{23}+k\epsilon_2
\end{equation}

\begin{equation}   \label{eq:301}
e^{'}_{33}=e_{33}+2k\epsilon_3-2z^2\left(c+3d\right)
\end{equation}

\begin{equation}   \label{eq:311}
e^{'}_{00}=e_{00}-2k\epsilon_0-2z^2\left(3a+b\right)
\end{equation}
 
the conditions $-a+b=\frac{\Lambda}{6}$ and $-c+6d+4e=\Lambda$ must be satisfied. Here we want to keep $e_{11}$ as a physically relevant component for the polarization tensor in agreement with \cite{1}. Then the following additional conditions must be imposed:

\begin{equation}   \label{eq:500}
c=-e\;\;\;-5c+6d=\Lambda
\end{equation}  
 
\begin{equation*}   
c=-3d\;\;\;d=\frac{\Lambda}{21}
\end{equation*} 

\begin{equation*}   
b=-3a\;\;\;a=-\frac{\Lambda}{24}
\end{equation*}

then the physical relevant components are:

\begin{equation}   \label{eq:2711}
e^{'}_{11}=e_{11}\;\;\;e^{'}_{12}=e_{12}
\end{equation}

the behavior under rotations is still \cite{1}:

\begin{equation}   \label{eq:33}
e^{'}_{\pm}=exp(\pm 2i\theta)e_{\pm}
\end{equation} 

with $e_{\pm}=e_{11}\mp i e_{12}$. This is what we were expecting. This result is in agreement with that obtained in \cite{S1} where it was found that $\Lambda$ does not affect the polarization of the GW during its propagation. In \cite{S1} $\Lambda$ only provides an isotropic contribution to the geodesic deviation equation given by:

\begin{equation}   \label{eq:55555}
\ddot{Z_1}=\ddot{Z_2}=\ddot{Z_3}=\frac{\Lambda}{3}Z_1
\end{equation}

Z is the geodesic deviation coordinate in agreement with \cite{S1}. Then if we have a group of particles forming a circle. They will move isotropically keeping the initial shape of the circle. In other words, the polarization is not affected by the presence of $\Lambda$.
This is valid for Minkowski, de-Sitter and Anti de-Sitter \cite{S1}. 

\section{Power and critical distance in the $\Lambda$ gauge}   \label{eq:s6}

In \cite{2}, the radiation flux of a gravitational wave when it propagates in an asymptotically de-Sitter space was calculated by taking $\Lambda$ as an additional source of radiation (de-Donder gauge). Here we want to explore if the same critical distance $L_{crit}$ can be obtained when we take $\Lambda$ as a gauge effect. We can perform the same calculations as in \cite{2}, but this time the first order scalar curvature is $R^{(1)}=0$ in agreement with the eqn. (\ref{eq:16}). The effective gravitational Poynting vector in the $\Lambda$ gauge is then given by:

\begin{equation}   \label{eq:34}
\hat{t}_{0 i}=\frac{1}{8\pi G}\left(R^{(2)}_{0 i}-\Lambda h_{0 i}\right)+O(h^3)
\end{equation}

This result differs from the one obtained in \cite{2} only in the first order contribution of the Ricci tensor $(R_{\mu \nu}^{(1)})$ which is zero in the present case. However, this difference is compensated by the gauge effect. 
If we replace the solutions given in eqn. (\ref{eq:20}), after some standard calculations and after making an average over a large region of spacetime, we obtain the same result $L_{crit}\approx (f\hat{h})r^2_\Lambda$ as that obtained in \cite{2}, repeated in eqn. (\ref{eq:15}) for clarity. The standard condition $r_{\Lambda}\approx \frac{\lambda}{h}$ \cite{7} when combined with the critical distance just found previously, implies:

\begin{equation}   \label{eq:35}
L_{crit}\approx r_{\Lambda}
\end{equation} 

and then the so called critical distance found originally in \cite{2} is just the background curvature scale \cite{7}. From an analysis of local physics we have just obtained the same global effect already explained in detail in section (\ref{eq:s1}) in order to explain the validity of the GW approach. Here however we interpret this result as a consequence of the power decay rate as the GW propagates inside the background. In other words, the background absorbs the energy of the wave as it propagates. This is in agreement with the CNC since it demonstrates the tendency for the inhomogeneities to be dissipated at large times (distances). It shows the tendency for the space to go asymptotically to the de-Sitter one.
In reference \cite{8K} the equations for the evolution of GWs in an asymptotically de-Sitter space were written as in the asymptotically flat case but including a viscosity term and a decaying term which is in complete agreement with the present formalism and interpretation. The formalism developed here is however very simple and deserves some attention. 

\section{A comparison with the Stochastic background of gravitational waves}   \label{eq:Stochastic}     

It is expected that the forthcoming interferometric experiments (VIRGO, LIGO, etc) will be able to detect some time in future the stochastic backgound of gravitational waves. We expect to find information about the origin of the universe from such observations. The intensity of the stochastic background GW is characterized by a dimensionless function of the frequency $\Omega_{GW}(f)\equiv\frac{1}{\rho_{crit}}\frac{d\rho_{GW}}{d lnf}$ \cite{B. Allen, Mic}. Where $\rho_c$ is the critical density necessary to close the universe and $\rho_{GW}$ is the energy density of the stochastic background GW. In terms of the present value of the Hubble constant, the critical density is given by $\rho_c=\frac{3H_0^2}{8\pi G}$. Where $H_0=h_0\times 100 Km/(sec-Mpc)$ is the actual value of the Hubble constant. $h_0$ parametrizes the existing experimental uncertainty. As $H_0$ is an uncertain quantity, this uncertainty is translated to the expression $\Omega_{GW}$ and for this reason the stochastic background of GWs is normally characterized by $h_0^2\Omega_{GW}(f)$. This background can also be characterized by other quantities as explained in refs. \cite{B. Allen, Mic}.     
\\
Different backgrounds provide different forms of the function $\Omega_{GW}$, and as a consequence different backgrounds will affect in different ways the propagation of gravitational waves. In \cite{C1, C2} for example, it is shown the behavior of GWs for different sources. Interesting is the case where the gravitational wave propagates inside a background dominated by a power-law inflationary phase. In such a case, the amplitude of the wave decreases as it propagates through the background (see the plots in \cite{C1}).\\
In \cite{C1} the background sources were introduced through a conformal transformation. In such a case, the amplitude $h$ remains unchanged and only the background, represented by the conformal factor, changes. In \cite{C2} the analysis of the propagation of GWs is performed by introducing an additional scalar component. Similar analysis were performed in \cite{B. Allen, Mic} inside a scalar-tensor theory. Here we just copy the results obtained for the case of de-Sitter inflationary phase scenario \cite{B. Allen, Mic, C2}. In such a case, the perturbation propagates through a FRW background with metric:

\begin{equation}   \label{eq:36}
ds^2=a^2(\eta)(-d\eta^2+dx^2)
\end{equation}  

where $d\eta=\frac{dt}{a(t)}$ is the conformal time. The propagation of a perturbation through the metric can be expressed as 

\begin{equation}
g_{\mu \nu}=a^2(\eta)(-d\eta^2+dx^2+h_{\mu \nu}dx^\mu dx^\nu) 
\end{equation}

If the GW perturbation is represented by:

\begin{equation}   \label{eq:37}
h_{\mu \nu}=e_{\mu \nu}\phi(\eta)e^{i\vec{k}\cdot\vec{x}}
\end{equation}

then the amplitude $\phi(\eta)$ has to satisfy the following equation:

\begin{equation}   \label{eq:38}
\left(\frac{d^2}{d\eta^2}+\frac{2}{a}\frac{da}{d\eta}\frac{d}{d\eta}+\vert\vec{k}\vert^2\right)\phi=0
\end{equation}

for the purposes of this paper, the relevant result is that corresponding to the de-Sitter inflationary phase. Where the scale factor behaves like $a\propto e^{H_{ds}t}$ in terms of the physical time $t$. For the de-Sitter scenario, the solution of eq. \ref{eq:38} becomes:

\begin{equation}   \label{eq:39}
\phi(\eta)=\frac{a(\eta_1)}{a(\eta)}\left(1+iH_{ds}\omega^{-1}\right)e^{-ik(\eta-\eta_1)}
\end{equation}

where $H_{ds}$ corresponds to the Hubble factor during the de-Sitter phase. From the result \ref{eq:39} it is clear that the amplitude has an oscillatory part but it suffers a damping (decreasing) effect due to the factor $\frac{a(\eta_1)}{a(\eta)}$. This factor in terms of the physical time $t$, is equivalent to an exponentially decreasing amplitude $\phi\propto e^{-H_{ds}t}$. This simply shows that inside a de-Sitter background the wave cannot propagate at a distance farther than $r\approx H_{ds}^{-1}$ (in geometrized units $c=1$). If we take the actual value for the Hubble scale $H_0\approx \frac{1}{r_\Lambda}$; then $H_0^{-1}$ plays the role of the critical distance found in eqs. \ref{eq:15} and \ref{eq:35} where the GWs were expanded around the Minkowski space but introducing $\Lambda$ as an additional source (de-Donder gauge) or as a gauge effect ($\Lambda$ gauge).

\end{document}